
%
%

 \hyphenation{Schro-ding-er}
 \def\lsb{[}      
 \def\rsb{]}      
 \def\xhat{^}     

 \mathsurround=1pt
 \magnification=\magstep1\openup 1\jot
 \raggedbottom
 
 \font\caps=cmcsc10
 
 \font\fn=cmr7  
 \def\square{{\vcenter{\vbox{\hrule height.4pt
         \hbox{\vrule width.4pt height6pt \kern6pt
         \vrule width.4pt}
         \hrule height.4pt}}}}
 \def\planck{\hbar}
 \def\prefactor{(a^4-k^2)^{-1/4}}

 \def\tendsto{\rightarrow}

 \def\ms{\hbox{\rm \hskip 10pt .}}
 \def\mc{\hbox{\rm \hskip 10pt ,}}
 
 \def\half{{1\over2}}

 \def\hh{\Psi_{H\!H}}

 \def\lm{{\ell m}}
 \def\nlm{{n\lm}}
 \def\QQ{Q\xhat n_\lm}
 
 \def\Sij{{\left(S_{ij}\right)}\xhat n_\lm}
 
 \def\Pij{{\left(P_{ij}\right)}\xhat n_\lm}
 \def\Gij{{\left(G_{ij}\right)}\xhat n_\lm}
 \def\Sn{\sum_n}

 \def\vac{\left.|0\right>}

 \def\eps{\epsilon}

 \newcount\sectionno \sectionno=0
 \newcount\eqnnumber \eqnnumber=0
 \newcount\subsectionno\subsectionno=1

 \def\appendix#1{

    \eqnnumber=0
    \def\nextno{
        \global\advance\eqnnumber by 1
        ({\rm #1}.\the\eqnnumber)
               }
 \def\eqlabel##1{
        \xdef##1{{$({\rm #1}.\the\eqnnumber)$}}
       }
 }

 \def\newsection#1{
    \vskip .5cm \goodbreak
    \global\advance\sectionno by 1
    \centerline{\bf #1}
    \nobreak\vskip .5cm\nobreak
    \eqnnumber=0 \subsectionno=1
 \def\nextno{
        \global\advance\eqnnumber by 1
        (\the\sectionno.\the\eqnnumber)
               }
 \def\eqlabel##1{
        \xdef##1{{$(\the\sectionno.\the\eqnnumber)$}}
       }
 \def\newsubsection##1{
    \vskip .5cm \goodbreak
    \global\advance\subsectionno by 1
    \eqnnumber=0
    \leftline{\bf\the\sectionno.\the\subsectionno\quad  ##1}
    \nobreak\vskip .5cm\nobreak
    \eqnnumber=0

 }}
 \def\eqnl{\eqno\nextno\eqlabel}
 \def\eqn{\eqno\nextno}

%
%
%
 \newwrite\refout
 \newcount\refno
 \def\refbegin{\immediate\openout\refout=refout.tex\refno=1}
 
 %
 %
 \def\form#1#2#3#4#5#6{\xdef#1{{\noexpand\rm#2,
  {\noexpand\rm#3\/}
   {\noexpand\bf#4} #6 (#5).}}}
 %
 %
 \def\bookform#1#2#3#4{\xdef#1{{\noexpand\rm#2, {\noexpand\it#3\/}
  (#4).}}}
 %
 %
 \def\procform#1#2#3#4#5{\xdef#1{{\noexpand\rm#2, in {\noexpand\it#3\/}
   (#4), (#5).}}}
 %
 %
 \def\prepform#1#2#3#4#5{\xdef#1{{\noexpand\rm#2, #3 (#4)#5.}}}
 %
 %
 \def\freeform#1#2{\xdef#1{{\noexpand\rm#2.}}}
 %
 %
 \def\refwrite#1{\immediate\write\refout{\noexpand#1}}
 \def\ref#1#2{\ifx#1\undefined\message{*Ref*}$\ast$\else
   \ifx#2\undefined\xdef#2{\number\refno}#2%
   \refwrite{\item{\lsb #2\rsb}\noexpand#1}%
   \global\advance\refno by 1\else#2\fi\fi}
 %
 %
 \def\xref#1#2{\ifx#1\undefined\message{*Ref*}$\ast$\else
   \ifx#2\undefined\xdef#2{\number\refno}%
   \refwrite{\item{\lsb #2\rsb}\noexpand#1}%
   \global\advance\refno by 1\else\fi\fi}
 %
 %
 \def\sref#1{\ifx#1\undefined\message{*Ref*}$\ast$\else
   \refwrite{\item{}\noexpand#1}\fi}
 %


 \procform{\swhhouches}{S. W. Hawking}{Relativity, Groups and Topology
    II, Les Houches 1983, Session XL}{B.~S.~DeWitt and R.~Stora}
    {North-Holland, Amsterdam, 1984}
\form{\hh}{J. B. Hartle and S. W. Hawking}{Phys. Rev.}{D28}{1983}{2960}
\form{\swhpl}{S. W. Hawking}{Phys. Lett.}{B195}{1987}{337}
\form{\swhpr}{S. W. Hawking}{Phys. Rev.}{D37}{1988}{904}
\form{\hl}{S. W. Hawking and R. Laflamme}{Phys. Lett.} {B209}{1988}{39}
\form{\halh}{J. J. Halliwell and S. W. Hawking}{Phys. Rev.}
{D31}{1985}{1777}
\form{\swhcc}{S. W. Hawking}{Phys. Lett.}{B134}{1984}{403}
\form{\swhdo} {S. W. Hawking}{Nucl. Phys.}{B335}{1990}{155}
\form{\hp}{S. W. Hawking and D. N. Page}{Phys. Rev.}{D42}{1990}{2655}
\form{\swhcmpetw} {S. W. Hawking}{Comm. Math. Phys.}{87}{1982}{395}
\procform{\swhgr} {S. W. Hawking}
{General Relativity: An Einstein Centenary Survey}{S. W. Hawking and W.
Israel}{Cambridge, 1979}
\form{\swhcmpsfi}{S. W. Hawking}{Comm. Math. Phys.}{43}
{1975} {199}
\form{\swhnew}{S. W. Hawking}{Nucl. Phys.}{B363}{1991}{117}
\form{\arlo}{A. Lyons}{Nucl. Phys.}{B324}{1989}{253}
\form{\arltw}{A. Lyons}{Nucl. Phys.}{B333}{1990}{279}
\form{\arlnew}{A. Lyons and S. W. Hawking}{Phys. Rev.}{D44}{1991}{3802}
\prepform{\arlun}{A. Lyons}{forthcoming}{1991}{}
\form{\hfd}{H. F. Dowker} {Nucl. Phys.}{B331}{1990}{194}
\form {\hfdl}{H. F. Dowker and R. Laflamme}{Nucl.
Phys.}{B366}{1991}{209}
\prepform{\hfdthesis}{H. F. Dowker}{PhD dissertation}{1990}{ (Cambridge
University)}
\form{\colbh}{S. Coleman}{Nucl. Phys.}{B307}{1988}{867}
\form{\colcc}{S. Coleman}{Nucl. Phys.}{B310}{1988}{643}
\form{\gsqc}{S. B. Giddings and A. Strominger}{Nucl.
Phys.}{B307}{1988}{854}
\form{\gsax}{S. B. Giddings and A. Strominger}{Nucl.
Phys.}{B306}{1988}{890}
\form{\lee}{K. Lee}{Phys. Rev. Lett.}{61}{1988} {263}
\form{\ksb}{I. Klebanov, L. Susskind and T. Banks}{Nucl.
Phys.}{B317}{1989}{665}
\form{\pope}{C. Pope}{J. Phys.} {A15}{1982}{2455}
\form{\dh}{P. D. D'Eath and J. J. Halliwell}{Phys. Rev.}{D35}
{1987}{1100}
\prepform{\dowpet}{J. S. Dowker and D. F. Pettengill}{}{unpublished}{
Spinor Hyperspherical Harmonics, Expansions on $SU(2)$
and some Applications}
\form{\jan}{R. T. Jantzen}{J. Math. Phys.}{19}{1978}{1163}
\form{\cas}{R. Casalbuoni}{Nuovo Cimento} {A33}{1976}{115}
\bookform{\dir}{P. A. M. Dirac}{Lectures on Quantum Mechanics}
{Academic, New
York, 1965}
\bookform{\ber}{F. A. Berezin}{The Method of Second Quantization}
{Academic, New
York, 1966}
\bookform{\ed}{A. R. Edmonds}{Angular Momentum in Quantum Mechanics}
{Princeton, 1957}
\form{\ggil}{G. Gilbert}{Nucl. Phys.}{B328}{1989}{159}
\form{\banks}{T. Banks}{Nucl. Phys.}{B309}{1988}{493}
\form{\fs}{W. Fischler and L. Susskind}{Phys. Lett.}{B173}{1986}{262}
\form{\adsfi}{M. Ademollo et al}{Nucl. Phys.}{B94} {1975}{221}
\form{\shap}{J. A. Shapiro}{Phys. Rev.}{D11}{1975} {2937}
\form{\polcc}{J. Polchinski}{Commun. Math. Phys.}{104}{1986} {37}
\form{\rohm}{R. Rohm}{Nucl. Phys.}{B237}{1984}{553}
\form{\iz}{C. Itzykson and J. B. Zuber}{Nucl. Phys.}{B275}{1986}{580}
\bookform{\gsw}{M. B. Green, J. H. Schwarz and E. Witten} {Superstring
theory}{Cambridge University Press, Cambridge, England, 1985}
\form{\adsfo}{M. Ademollo et al}{Nucl. Phys.}{B77}{1974}{189}
\form{\ns}{A. Neveu and J. Scherk}{Nucl. Phys.}{B36} {1972}{317}
\form{\es}{U. Ellwanger and M. G. Schmidt}{Z. Phys.}{C43} {1989}{485}
\form{\os}{H. Ooguri and N. Sakai}{Nucl. Phys.}{B312}{1989}{435}
\form{\mis}{C. W. Misner}{Phys. Rev.}{118}{1960}{1110}
\form{\gh}{G. W. Gibbons and S. W. Hawking}{Phys. Rev.}{D15}
{1977}{2752}
\form{\polphase} {J. Polchinski}{Phys. Lett.}{B219}{1989}{251}
\form{\polfac}{J. Polchinski}{Nucl. Phys.}{B307}{1988}{61}
\form{\fswc} {W. Fischler and L. Susskind}{Phys. Lett.}{B217}{1989}{48}
\form{\polret}{J. Polchinski}{Nucl. Phys.}{B325}{1989}{619}
\form{\poly}{A. M. Polyakov}{Phys. Lett.}{B103}{1981}{207}
\form{\sw}{N. Seiberg and E. Witten}{Nucl. Phys.}{B276}{1986}{272}
\procform{\man}{S. Mandelstam}{Unified String Theories, Proceedings of
the Santa Barbara Workshop,}{M. Green and D. Gross}{World Scientific,
Singapore, 1986}
\form {\pres} {J. Preskill}{Nucl. Phys.}{B323} {1989} {141}
\form{\hall} {J. J. Halliwell and R. Laflamme}{Class. Quant. Grav.}{6}
{1989}{1839}
\form{\ho}{A. Hosoya and W. Ogura}{Phys. Lett.}{B225}{1989}{117}
\form {\gw} {B. Grinstein and M. B. Wise}{Phys. Lett.}{B212} {1988}
{407}
\form {\gp} {D. J. Gross and V. Periwal}{Phys. Rev. Lett.} {60} {1988}
{2105}
\form{\gpope}{G. W. Gibbons and C. N. Pope}{Commun. Math.
Phys.}{66}{1979}{267}
\bookform{\gj} {J. Glimm and A. Jaffe}{Quantum physics, a functional
integral
point of view} {Springer, 1981}
\form{\gero}{R. P. Geroch}{J. Math. Phys.} {8} {1967} {782}
\form{\gertw}{R. P. Geroch}{J. Math. Phys.} {11} {1970} {437}
\form{\ghpw}{A. K. Gupta, J. Hughes, J. Preskill and M. B. Wise}{Nucl.
Phys.}{B333}{1990}{195}
\form{\dewitt}{B. S. DeWitt}{Phys. Rev.}{160} {1967} {1113}
\procform{\wheeler}{J. A. Wheeler}{Battelle
Rencontres}{C. DeWitt and J. A. Wheeler} {Benjamin, New York, 1968}
\form{\coll}{S. Coleman and K. Lee}{Phys. Lett.}{B221}{1989}{242}
\form{\gsth}{S. B. Giddings and A. Strominger}{Nucl.
Phys.}{B321}{1989}{481}
\form{\cmnp}{A. Cohen, G. Moore, P. Nelson and J. Polchinski}{Nucl.
Phys.}{B281}{1987}{127}
\form{\fks}{W. Fischler, I. Klebanov and L. Susskind}{Nucl.
Phys}{B306}{1988}{271}
\form{\sd}{S. Davis}{Class. Quant. Grav.}{6}{1989}{1791}
\form{\gm}{B. Grinstein and J. Maharana}{Nucl. Phys.}{B333}{1990}{160}
\form{\gms}{B. Grinstein, J. Maharana and D. Sudarsky}{Nucl.
Phys.}{B345}{1990}{231}
\prepform{\aato}{A. A. Tseytlin}{IC/89/90}{1989}{ Sigma models and
renormalization of string loops, Lectures given at the 1989 Trieste
Spring School on Superstrings, ICTP-Trieste preprint}
\form{\aattw}{A. A. Tseytlin}{Phys. Lett.}{B251}{1990}{530}
\form{\rnso}{P. Ramond}{Phys. Rev.}{D3}{1971}{2415}
\form{\rnstw}{A. Neveu and J. H. Schwarz}{Nucl. Phys.}{B31}{1971}{86}
\form{\gsoo}{F. Gliozzi, J. Scherk and D. Olive}{Phys.
Lett.}{B65}{1976}{282}
\form{\gsotw}{F. Gliozzi, J. Scherk and D. Olive}{Nucl.
Phys.}{B122}{1977}{253}
\form{\red}{A. N. Redlich}{Nucl. Phys.}{B304}{1988}{129}
\form{\alv}{O. Alvarez}{Nucl. Phys.}{B216}{1983}{125}
\bookform{\riemannsurfaces}{J. D. Fay}{Theta Functions on Riemann
Surfaces, \noexpand{\rm Springer Notes in Mathematics 352}}{Springer,
1973}
\form{\polver}{J. Polchinski}{Nucl. Phys.}{B289}{1987}{465}
\bookform{\arf}{G. Arfken}{Mathematical Methods for
Physicists}{Academic, New
York, 1970}
\form{\ghp}{G. W. Gibbons, S. W. Hawking and M. J. Perry}{Nucl.
Phys.}{B138}{1978}{141}
\form{\dhp}{E. D'Hoker and D. H. Phong}{Rev. Mod. Phys.}{60}{1988}{917}
\form{\jjhcon}{J. J. Halliwell and J. B. Hartle}{Phys.
Rev.}{D41}{1990}{1815}
\form{\hpold}{S. W. Hawking and D. N. Page}{Nucl.
Phys.}{B264}{1986}{185}
\bookform{\feyn}{R. P.  Feynman and A. R. Hibbs}{Quantum
Mechanics and Path integrals}{McGraw-Hill, 1965}
\form{\jjhjl}{J. J. Halliwell and J. Louko}{Phys. Rev.}{D40}{1989}{1868}
\procform{\adm}{R. Arnowitt, S. Deser and
C. Misner}{Gravitation: An Introduction to Current Research}{L.
Witten}{Wiley, New York, 1962}
\form{\poltoy}{J. Polchinski}{Nucl. Phys.}{B324}{1989}{123}
\bookform{\titchft}{E. C. Titchmarsh}{Introduction to the theory of
Fourier integrals}{Oxford, Clarendon Press, 1937}
\prepform{\arlwip}{A. Lyons}{work in preparation}{1992}{}
\bookform{\bateman}{A. Erdelyi, W. Magnus, F. Oberhettinger and F.
Tricomi (eds.)}{Bateman
Manuscript Project: Higher Transcendental Functions \noexpand{\rm Vol.
2}}{McGraw Hill, New York, 1953}
\form{\ll}{R. Laflamme and J. Louko}{Phys. Rev.}{D43}{1991}{3317}
\prepform{\gundlach}{C. Gundlach}{Gauge invariant quantizations of
cosmological perturbations}{1991}{ DAMTP preprint}
\form{\wada}{S. Wada}{Phys. Rev.}{D34}{1986}{2272}
\form{\shw}{I. Shirai and S. Wada}{Nucl. Phys.}{B303}{1988}{728}
\form{\sz}{E. Salusti and F. Zirilli}{Nuovo Cimento Lett.}{4}{1970}{999}
\form{\berg}{B. K. Berger}{Phys. Rev.}{D11}{1975}{2770}

\nopagenumbers
\hbox{ }
\rightline{ALBERTA THY-3-92}
\bigskip\bigskip
{\bf
\centerline{ALL-ORDERS WORMHOLE VERTEX OPERATORS}
\smallskip
\centerline{FROM THE WHEELER-DEWITT EQUATION}
}
\vskip .6cm
\centerline{ALEX LYONS\footnote*{\fn Electronic mail:
ALYONS@FERMI.PHYS.UALBERTA.CA}}
\vskip .3cm
{\bf
\centerline{Theoretical Physics Institute, Department of Physics}
\centerline{University of Alberta}
\it
\centerline{Edmonton, Alberta, T6G 2J1, CANADA}
}\vskip .5cm
\vskip.5cm
{\centerline{Submitted to Nucl. Phys. B, Jan 1992}}
\vskip .5cm
{\bf \centerline{Abstract}}
\vskip .5cm

{\leftskip 10truemm \rightskip 10truemm

\openup -1\jot

\noindent
We discuss the calculation of semi-classical wormhole vertex operators
from wave functions
which satisfy the Wheeler-deWitt equation and momentum constraints,
together with certain
`wormhole boundary conditions'.
We consider a massless minimally coupled scalar field, initially in the
spherically
symmetric `mini-superspace' approximation, and then in the
`midi-superspace'
approximation, where non-spherically symmetric perturbations are
linearized about a
spherically symmetric mini-superspace background. Our approach suggests
that there are
higher derivative corrections to the vertex operator from the
non-spherically symmetric
perturbations. This is compared directly with the approach based on
complete wormhole
solutions to the equations of motion where it has been claimed that the
semi-classical
vertex operator is exactly given by the lowest order term, to all orders
in the size of
the wormhole throat. Our results are also compared with the conformally
coupled case.

\openup 1\jot

}

\vfill\eject
\pageno=1
\footline={\hss\tenrm\folio\hss}

\refbegin
\newsection{1 Introduction}

Wormholes can be thought of as euclidean solutions to the field
equations for
gravity, possibly coupled to some matter system, which connect two
asymptotically flat
regions.\footnote*{\fn By `asymptotically flat' we mean `asymptotically
euclidean in the
sense of Gibbons and Pope [\ref\gpope\GPOPE]'.}
These solutions can be thought of as contributing
to the semi-classical (zero-loop) approximation to the partition
function for
gravity, and thereby to the Green functions for fields in the asymptotic
regions. There
are, however, problems with just considering real euclidean solutions.
Firstly, it is
known that a complex contour is needed in order to make the path
integral for gravity well
defined, and it is not clear how that contour should be defined,
particularly when matter
fields are present. Thus it would seem that arbitrary complex
saddle-points, in which both
geometry and matter are complex, might be just as relevant to the
semi-classical
evaluation of the partition function as real euclidean solutions.
Secondly, it is known
that in order for real euclidean wormhole solutions to exist, one needs
very particular
types of matter fields, such as a conformally coupled, or imaginary
minimally coupled,
scalar field. However, for wormholes to provide a viable solution to the
cosmological
constant problem [\ref\colcc\COLCC], or to be relevant to the late
stages of black hole
evaporation [\ref\swhpr\SWHPR],
one would like their existence to be independent of the particular
matter fields in the system.

It is for these reasons that Hawking and Page [\ref\hp\HP] introduced
the idea that
wormholes should be considered instead as solutions to the
Wheeler-deWitt equation. The
wave functions of wormholes would satisfy particular boundary conditions
corresponding to
the fact that they describe states which, semi-classically, correspond
to four-geometries
connecting two asymptotically flat regions. Hawking and Page show that
one can find a
complete spectrum of such wave functions which generalizes to a rather
large class of
matter field models. However, they did not extract a low-energy
effective action from
their wave functions. It is generally expected that at low energies it
is possible to
replace the wormhole ends by vertex operators in the effective field
theory, which would
then give rise to an effective bilocal action. The aim of this paper is
to show that it is
possible to extract the vertex operators directly from the wave
functions. The approach
based on wormhole wave functions turns out to be more general than using
euclidean
solutions to the field equations, in that it potentially allows a wider
class of operators
in the effective action. In general we would expect the effective action
to include all
possible gauge-invariant and Lorentz-covariant operators. The euclidean
solution approach
fails to find all these operators, due to the restricted nature of
possible solutions. For
example, in the imaginary scalar field model, the euclidean wormhole
solutions have to be
$O(4)$-symmetric. Grinstein, Maharana and Sudarsky have calculated the
semi-classical
vertex operator corresponding to these wormholes, to all orders in the
wormhole throat
radius [\ref\gms\GMS], and they show that there are no derivative terms
in the effective
action. However, other work based on the wormhole wave functions
suggests that one obtains
these derivative terms [\ref\swhpr\SWHPR,
\ref\arlo\ARLO--\xref\hfd\HFD\ref\hfdl\HFDL].

We would like to point out that considering just solutions to the
Wheeler-deWitt equation
is, itself, probably an approximation. It is now realized (using a
simple two-dimensional
model) [\ref\arlnew\ARLNEW] that the wormhole wave functions need not
actually be
`on-shell', i.e. satisfying the Wheeler-deWitt equation. In the
two-dimensional model
considered in [\ref\arlnew\ARLNEW], the `on-shell' states only provide
the dominant
contribution in the situation where the wormholes are long and thin. The
`on-shell' states
appear as poles in the vertex operator, when the integration over the
size of the wormhole
throat is performed. In four-dimensional gravity we do not know the
measure in the
integration over the size of the wormhole. However, we shall assume that
it is such that
`on-shell' states provide the dominant contribution to the low-energy
effective action, as
in the two-dimensional model.

The plan of this paper is as follows: In Sec.~2, we consider the
spatially homogeneous,
$O(4)$-symmetric, massless scalar field model. We show how the $\colon
e^{ik\phi}\colon$
effective interactions, usually associated with the euclidean axionic
(or imaginary scalar
field) wormhole solutions, can be thought of as arising from a basis of
solutions to the
Wheeler-deWitt equation. Thus the wave function approach (in the
$O(4)$-symmetric mini-superspace
truncation) reproduces the results usually obtained by
considering euclidean
wormhole solutions. In Sec.~3, we discuss the non-spherically
symmetric perturbations of the
axion wormhole in the wave function approach. We
treat the problem in the `midi-superspace approximation'
as used in [\ref\halh\HALH]. Excitations of the lowest spatially
inhomogeneous mode
($n=2$) would be expected to produce a tower of effective interactions
of the form
$\colon(\partial\phi)^m\,e^{ik\phi}\colon$, for even $m$. This is based
on the fall-off
behaviour of the modes in the asymptotic region. However, a more careful
analysis of the
linearized lapse and shift constraints shows that only the ground state
wave function is
allowed. This leads to just the vertex operator with $m=0$.
In contrast, for the higher spatially
inhomogeneous scalar modes one obtains higher derivative corrections
to the vertex operator.
Our findings for the lowest mode are in agreement with the
`non-renormalization
theorem' of Grinstein et al [\ref\gms\GMS]. However, it is claimed in
[\ref\gms\GMS] that
the vertex operator $\colon e^{ik\phi}\colon$ is correct to {\it all}
orders in the size
of the wormhole. In our formalism, one would expect higher derivative
corrections to this
operator. We discuss the differences between our conclusions and those
of [\ref\gms\GMS]
and also compare our results with those of Dowker
[\ref\hfdthesis\HFDTHESIS] in the conformally coupled case.
We summarize our main results in Sec.~4. Finally,
in the Appendix, we discuss in more detail the construction of the
solutions to the euclidean Schrodinger equation which are used in
Sec.~3.

\newsection
{2 Wormholes, scalar fields, and the Wheeler-deWitt equation}

The aim of this section is to relate the wave function description of
wormholes
to the more usual description in terms of real euclidean wormhole
solutions.
We start by considering the simplest mini-superspace model which
exhibits wormhole
solutions. This is the model consisting of gravity
minimally coupled to a massless scalar field. The scalar field can be
thought of as the
Goldstone boson corresponding to a spontaneously broken global $U(1)$
symmetry [\ref\coll\COLL]. Alternatively there is a dual description of
this model
(related by a simultaneous canonical and Hodge duality transformation)
in terms
of an anti-symmetric three-index field strength. It was in this form
that wormhole
solutions were first found [\ref\gsax\GSAX]. However, we shall here (for
simplicity) work
with the scalar field formulation.
In this model the spatial sections are
 taken to be three-spheres and the metric ansatz takes the form:
 $$ds^2=\sigma^2\left(N^2(t)dt^2+a^2(t)d\Omega_3^2\right)\mc\eqn$$ with
the
matter field taken to be homogeneous on the spatial sections.
The constant $\sigma^2=2/3\pi m_p^2$ is included for later convenience,
and
$d\Omega_3^2$ is the usual round metric on the unit three-sphere.

 With this ansatz the euclidean action for this mini-superspace model is
 $$I=-\half\int dt\,Na^3\left({1\over N^2a^2}\left(da\over dt\right)^2
 +{1\over a^2} - {1\over N^2}\left(d\phi\over dt\right)^2\right) +
 \half\left(a(t_-)\right)^2\ms\eqnl\action$$
 This is the action appropriate for the boundary value problem where $a$
and
$\phi$ are fixed on the two boundaries, labelled by $t_\pm$. The
boundary term
 in the action is present in the case where the boundary labelled by
$t_-$ is
taken to become asymptotically flat space. This is the case which we
shall be
interested in throughout this paper. The limit $t_-\tendsto -\infty$ is
the
asymptotically flat region of the space-time.
 Varying the action with respect to $N$ and $\phi$ leads to the
constraint
equation (the hamiltonian constraint for gravity) and the equation of
motion for $\phi$
which are:
 $$\eqalign{
 {1\over N^2}\left(da\over dt\right)^2&=1+{a^2\over N^2}\left(d\phi\over
dt\right)^2\cr
 {d\over dt}\left({a^3\over N}{d\phi\over
dt}\right)&=0\ms}\eqnl\eqofmot$$
 These form a complete set of equations for the system, since if these
hold at all times then
 the $a$ equation
 of motion is automatically satisfied.

 The simplest general form of the solution to these equations can be
obtained
in the gauge where $N=1/a$. In this case
 \eqofmot\ can be integrated trivially
 to yield the solution
 $$\eqalign{
 ds^2&=\sigma^2\left(a^{-2}(t)dt^2+a^2(t)d\Omega_3^2\right)\cr
 a^2&=\sqrt{4t^2-Q^2}\cr
 \phi-\phi_-&={1\over4}\ln\left(2t-Q\over
2t+Q\right)\mc}\eqnl\solution$$
 where the constant of the motion (the euclidean momentum conjugate to
$\phi$) is
$$Q=a^4{d\phi\over dt}=a^2\sinh
 2(\phi-\phi_-)\ms\eqn$$
Here $\phi_-$ is the asymptotic value of $\phi$ at infinity, and a
constant
 of integration can also be added to $t$.

Flat space is recovered as the special case $Q=0$, while the wormhole
solution is
found by taking $Q$ to be purely imaginary. The metric then has two
asymptotically
flat regions, as $t\tendsto\pm\infty$, with a throat at $t=0$. This
solution
will also have a purely imaginary $\dot\phi$. One might argue that this
solution is
a bit of a cheat, since the matter field has negative definite energy
density and
seems rather unphysical. However, there are two good reasons for arguing
that such
a configuration may have relevance. Firstly, from the point of view of
providing
a semi-classical contribution to the partition function, it is known
that the
contour of integration must be distorted into the complex plane for
convergence
of the path integral. In that case one might argue that any saddle-point
may
be important, even one in which the metric and matter field become
complex
[\ref\jjhjl\JJHJL]. The
second argument comes from viewing the semi-wormhole solution
($-\infty<t\le0$)
as providing a semi-classical contribution to the amplitude for a
tunnelling
process in quantum gravity. The idea is that if one wants the amplitude
to tunnel
between states of definite real {\it lorentzian} momentum for the scalar
field,
then the boundary data ensures that $Q$ must be purely imaginary, as in
this
solution [\ref\lee\LEE].

In the alternative treatment of wormholes which uses the Wheeler-deWitt
equation,
we can bypass the argument over whether or not one should consider
complex
solutions. One might expect that a WKB approximation to a solution
of the Wheeler-deWitt equation could be written in the form
$e^{-I_{sp}(a,\phi)}$, where $I_{sp}(a,\phi)$ is the euclidean action of
the
classical solution \solution\ between asymptotically flat space as
$t\tendsto-
\infty$, and an inner boundary given by $t_+$, on which $a$ and $\phi$
are
fixed {\it and real}.
The action of this classical solution, given by substituting \solution\
into
\action, is  $$I_{sp}(a,\phi)=-t_+=\half
a^2\cosh2(\phi-\phi_-)\mc\eqnl\spaction$$
 which reproduces the flat space action
 in the case $\phi=\phi_-$. Notice that the solution which we use has
real
$\phi$. We do not need to go to complex $\phi$, because we do not need
to find
a classical solution which has two asymptotically flat regions. Instead,
we are
 interested in
the wave function for real values of its arguments.
 The Wheeler-deWitt equation for this model is
 $${1\over a}{\partial\over\partial a} \left(a{\partial\psi\over\partial
a}\right) - {1\over a^2}{\partial^2\psi\over\partial\phi^2} - a^2\psi =
0\mc\eqn$$ where the factor ordering has been chosen to be the one which
is
covariant under general co-ordinate transformations in the $(a,\phi)$
mini-superspace. The family of wave functions
 $$\psi_{\phi_-}(a,\phi)=e^{-\half a^2\cosh2(\phi-\phi_-)}\eqn$$ in fact
solve this
equation exactly,
although the derivation from the
euclidean action of a classical solution
 suggests that one would have only expected to satisfy this equation to
 leading order in the WKB approximation.

The following questions now arise: Firstly, in what way are the wave
functions $\psi_{\phi_-}$
related to the euclidean wormhole solution to the field
equations with an imaginary scalar field? If there is a straightforward
connection between them, one can then ask whether
the low-energy vertex operators for these wormholes can be obtained
directly from the wave functions. In the past vertex operators
have been obtained by considering Green functions on the
euclidean background solution [\ref\gm\GM]. One motivation for
using the wave functions instead comes
from the possibility of generalizing this model
to include spatially inhomogeneous perturbations. One
expects that these extra degrees of freedom would enlarge the space
of wormhole quantum states.
The vertex operators corresponding to these extra states would
perhaps be missed in a treatment which just considers the field theory
on a background $O(4)$ symmetric solution. These
will be the questions we shall consider in this paper.

 To address the first question, we first consider the Fourier transform
(with
respect to $\phi$) of our wave function. In other words we consider
$$\hat\psi_{\phi_-
}(a,k)=\int_{-\infty}^\infty d\phi\,e^{ik\phi}\,\psi_{\phi_-
}(a,\phi)\ms\eqnl\ftwf$$
 This can be thought of as changing from the $(a,\phi)$ to the $(a,k)$
representation of the state $\left.|\phi_-\right>$.
$\hat\psi$ is given explicitly by
a modified Bessel
function of imaginary order, $K_{\half ik}(a^2/2)$, multiplied by
$e^{ik\phi_-}$. It has
the approximately exponential form
[\ref\bateman\BATEMAN] $$\hat\psi\sim\prefactor\exp\left[-\half\left(
 \sqrt{a^4-k^2}+k\arcsin(k/a^2)\right)+ik\phi_-\right]\mc\eqn$$
in the region $a^4\gg k^2$ and $k^2\gg1$.

This form is suggestive of a euclidean saddle-point approximation to a
path integral
representation for $\hat\psi$. Indeed, consider the complex
canonical transformation
from $(a(t),\phi(t))$ to $(a(t),k(t))$, where $k=-ia^3\dot\phi/N$.
For $a^4>k^2$, the classical
variational problem with $(a,k)$ fixed on the inner boundary
 and $\phi\tendsto\phi_-$ at infinity
has a euclidean stationary point whose action is precisely
the negative of the
value in the exponent above. The action \action\ has to be supplemented
by a matter field
dependent boundary term on the inner boundary given by
$-ik(t_+)\phi(t_+)$, since the
classical variational problem has changed from one in which $\phi$ is
fixed on the
boundary to one in which $k$ is fixed. This boundary term is fixed by
the mathematical
requirement that the variational principle subject to the new boundary
conditions should
yield the equations of motion. It is this boundary term which yields the
$\half k\arcsin
(k/a^2)-ik\phi_-$ term in the evaluated euclidean action. The
$\half\sqrt{a^4-k^2}$ term
comes from \action\ evaluated at the stationary point.

The saddle-point four-geometry in the $(a,k)$ boundary value problem is
precisely a part
of the imaginary scalar field wormhole outside a three-sphere of given
radius $\sigma a$,
carrying `charge' $k$. (This would be the lorentzian $U(1)$ charge
carried by the wormhole
in the Goldstone boson interpretation of the theory.) The boundary value
problem has two
solutions however, unlike in the case where the field $\phi$ is
specified on the boundary.
These correspond to taking more than or less than half of the wormhole.
The evaluated
action for each solution differs only in the sign taken for the square
root in
$\half\sqrt{a^4-k^2}$. The solution which is related to the Fourier
transform of the wave
function $e^{-\half a^2\cosh2(\phi-\phi_-)}$ corresponds to taking less
than half the
wormhole, and the positive square root. This means that the wave
function goes like $e^{-
a^2/2}$ as $a\tendsto\infty$. The $(a,\phi)$ representation has the
advantage that the
path integral appears to pick out the semi-classical wave function
uniquely whereas the
$(a,k)$ representation does not. This is similar to the situation
encountered by Hartle
and Hawking in defining the semi-classical approximation to the wave
function of the
universe in pure Einstein gravity with a cosmological constant
[\ref\hh\HH]. In that case,
there are two euclidean four-geometries satisfying the field equations
with a three-sphere
boundary of given radius $\sigma a$ and no other boundary, but if
instead the conjugate
momentum to $a$ is specified on the boundary, then there is only one.

The analysis above shows a connection between the Fourier transformed
wave function
$\hat\psi$ and the imaginary scalar field wormholes. We see from it that
it is natural to
take the boundary condition that the wave functions fall off like
$e^{-a^2/2}$ at large
$a$. These wave functions correspond to saddle-point evaluations of path
integrals where
the four-geometries are taken to be asymptotically flat. We obtained a
representation of the states $\left.|\phi_-\right>$
in terms of either $(a,\phi)$ or $(a,k)$. The $(a,k)$
representation of $\left.|\phi_-\right>$
can be related to imaginary scalar field wormhole solutions of
the euclidean field equations. The $(a,\phi)$ representation suggested
the correct boundary conditions for wormhole wave functions
at large $a$. The next step is to ask whether it is
possible to use the
wave functions we have found to evaluate the vertex operators, and thus
the effective
action, corresponding to the euclidean wormhole solutions. We proceed by
analogy with the
conformally coupled scalar field [\ref\swhpr\SWHPR] and the case of
massless fermions,
photons or gravitons [\ref\arlo\ARLO\xref\hfd\HFD--\ref\hfdl\HFDL].
Namely, we assume that
there is a Hilbert space spanned by wormhole quantum states,
$\left.|\phi_-\right>$. The states $\left.|k\right>=\int d\phi_-\,
e^{-ik\phi_-}
\left.|\phi_-\right>$ also span this space. Each state has a vertex
operator associated to it, which reproduces the effect of the state
on low-energy $n$-point functions in the asymptotic region.
One can try to calculate the vertex operators corresponding to the
basis $\left.|\phi_-\right>$, where one fixes $\phi\tendsto\phi_-$
at infinity. But the vertex operators turn out to be especially simple
if the basis $\left.|k\right>$
is used. In the $(a,\phi)$ representation the wave functions
for these states are `plane wave' products, of the form
$$\psi_k(a,\phi)=K_{-\half ik}\bigl(a^2/2\bigr)\,
e^{-ik\phi}\ms\eqnl\plwaveprod$$
These wave functions satisfy the Wheeler-deWitt equation and are
exponentially damped at
large $a$. This boundary condition is suggested
by the previous discussion of the saddle-point approximation
to a path integral representation of the relevant wave functions. We
are, however, being
more general than [\ref\hp\HP] in our choice of boundary conditions in
that we do not
require any particular regularity condition at small $a$. Our view is
that the Wheeler-deWitt equation
is only an effective theory of quantum gravity, which
ceases to be
strictly valid at small $a$. (The semi-classical approximation breaks
down there.) Thus
any boundary conditions at small $a$ should come from a more fundamental
theory of quantum gravity, and not be imposed without justification.

The effect of these wormhole states on the field theory $n$-point
functions is given by the matrix element
$$\left<k|\phi(x_1)\cdots\phi(x_n)|0\right>\mc\eqnl\matrix$$ as in
[\ref\swhpr\SWHPR]. This is essentially the dilute wormhole
approximation, in
that we are considering each wormhole independently and assuming that
Green
functions with points in the two asymptotically flat regions factorize.
(We suspect that wave functions would have to be replaced by density
matrices in a
description which goes beyond the dilute wormhole approximation
[\ref\arlwip\ARLWIP].) The vacuum state $\vac$ is given by
$\left.|\phi_-=0\right>$.
The matrix element has a path integral representation as
 $$\int da_0 \mu(a_0)\,
 \int d\phi_0 \, \bar\psi_k(a_0,\phi_0)
 \int[dg][d\phi] \,
 e^{-I[g,\phi]} \, \phi(x_1) \cdots \phi(x_n) \mc\eqn$$
 where the inner path integral is taken over asymptotically flat
 four-geometries with an inner boundary on which $a=a_0$ and
$\phi=\phi_0$,
and with $\phi\tendsto0$ at infinity.  We integrate in the range
$0<a_0<\infty,\ -\infty<\phi_0<\infty$ with some unknown measure
$\mu(a_0)$. The saddle-point
approximation, applied to the inner integral, produces the
result $$\int d^4x_0\,\int da_0\Delta(a_0)\,
\int d\phi_0\,\bar\psi_k(a_0,\phi_0)\,
 e^{-\half a_0^2\cosh2\phi_0}\, \biggl(\half a_0^2\sinh2\phi_0\biggr)^n
\,\prod_{j=1}^n(x_j-x_0)^{-2}\mc\eqnl\spme$$ for the matrix element,
where the new co-ordinates $x$ are related to the old $t$ co-ordinate by
$$(x-x_0)^2=-
2t\mc\eqn$$
and we use \solution--\spaction. The co-ordinates $x^\mu$ become
the usual Cartesian co-ordinates in the asymptotically flat region, and
$x_0$ can be
interpreted as the position of the wormhole end in asymptotically flat
space. To obtain
\spme, we have also expanded the logarithm which occurs in the
saddle-point solution
\solution\ for $\phi$, in the asymptotic region
$t_j\tendsto-\infty$. The prefactor $\Delta(a_0)$ comes from the
determinant of the
fluctuations about the saddle-point, and from $\mu(a_0)$.
Its exact form need not concern us, and is irrelevant in the leading
semi-classical
approximation. The scalar field action being purely quadratic, we know
that $\Delta$ is
just a function of $a_0$. The variables $x_0$ are the zero-mode
co-ordinates. They must be
integrated over at the end of the calculation. In the simple model in
which $\phi$ is
homogeneous on the three-sphere spatial sections there is no additional
integration over
angular zero modes although in a more general model these integrals
would also be present.

 Now the integral over $\phi_0$ may also be approximated using the
saddle-point
 method. (This is an ordinary integral over $\phi_0$ in the spatially
 homogeneous model.) The saddle-point values for $\phi_0$ are the roots
of $$ik=a^2\sinh2\phi_{0\,sp}\mc\eqn$$
 so that in this approximation each of the $n$ factors of $\half
a_0^2\sinh2\phi_0$ are replaced by their
saddle-point values, $\half ik$. (There is also
another prefactor, which depends on $k$ and $a_0$,
 but not on $n$.) The $a_0$ integral gives a factor which will
depend on $k$ but not on $n$. We shall therefore ignore it, as it can be
absorbed in the normalization of the vertex operator.
 The matrix element in the semi-classical limit (and for $|x_j-
x_0|\tendsto\infty$)
therefore assumes the form
 $$\int d^4x_0 \,\alpha(k)\left(\half ik\right)^n\prod_{j=1}^n(x_j-
x_0)^{-2}\ms\eqn$$

This can be identified with a flat space correlation function
$\left<V^k_n \phi(x_1)\cdots\phi(x_n)\right>$,
provided the operator $V^k_n$ is taken to be
$$V^k_n=\int d^4x_0\,{\alpha(k)\over n!}\,
\colon(ik\phi(x_0))^n\colon\ms\eqnl\vn$$ Thus we obtain that
 the semi-classical effect of the wave function $\psi_k(a,\phi)$ on
$n$-point
 functions in the asymptotic region is the same as the insertion of the
 operator $V^k_n$ in the flat space $n$-point functions, for some
function $\alpha(k)$.
Crucial to this identification is the fact that the flat space $\phi$
propagator, $\left<\phi(x)\phi(0)\right>=\half x^{-2}$,
falls off with the same power of proper distance
in the asymptotic region as the saddle-point
solution for $\phi(x)$, given by \solution\ with $\phi_-=0$.
The factor of $n!$ in \vn\ is present because each of the $n$ factors in
the vertex operator can be contracted with any of the $n$ fields in the
asymptotic region, and there are $n!$ ways of doing this.
The vertex operator we have found depends on $n$, but we require a
vertex operator which
reproduces the matrix element for {\it any} $n$.
It must take the form of a sum $$V^k=\sum_0^\infty V^k_n= \int d^4x_0
\,\alpha(k)\,
\colon e^{ik\phi(x_0)}\colon\ms\eqnl\vertexop$$
The effect of this operator on $n$-point functions is the same as \vn,
to lowest order in $\planck$, which can be reinstated by
multiplying each contraction of $\phi$'s
by $\planck$, and dividing $ik\phi$ in the vertex operator by $\planck$.
The
correct result to lowest order in $\planck$ is all we can expect to
obtain,
since we have used the saddle-point approximation throughout.
At this point we would like to remark that this analysis
has just taken into account a
single wormhole end. The inclusion of an arbitrary number
of wormholes will lead to an effective integral over the coupling
constants
$\alpha(k)$, for each value of $k$.
This is similar to a functional integral over the field $\alpha(k)$ on
superspace. $\alpha$ becomes a kind of quantized field on superspace
[\ref\arlnew\ARLNEW].
Whether there is some mechanism such as `the Big Fix' [\ref\colcc\COLCC]
to fix the values
of $\alpha(k)$ uniquely, remains unclear.

The vertex operator \vertexop\ is exactly what was found in
[\ref\gm\GM],
which uses the imaginary scalar field wormhole solution. However, it is
important that nowhere here is it necessary to assume the existence of a
complete wormhole
solution interpolating between two asymptotically flat regions. We
believe that, being not
so reliant on exact wormhole solutions to the classical field equations,
this derivation
of the effective interaction will generalize more readily to more
realistic models, such
as those in which the restriction of homogeneity is removed. We shall
now go on to discuss
these more general models.

\newsection{3 Spatially inhomogeneous perturbations}

It has been suggested by Grinstein et al [\ref\gms\GMS] that the vertex
operator for
massless scalar field wormholes given by \vertexop\ is correct,
to all orders in $l_{pl}$, the
Planck length, and to lowest order in $\planck$. The calculations of
[\ref\gms\GMS] are
based on the Green functions of test fields in an
$O(4)$-symmetric euclidean wormhole background.
We shall see in
this section that to obtain derivative effective interactions (which
will be higher order
in $l_{pl}$), it is necessary to consider inhomogeneities on the
three-sphere sections.
There will not be a complete wormhole solution connecting two
asymptotically flat regions,
but in the wave function approach this does not matter. What is
important is that there is
a family of solutions to the Wheeler-deWitt equation with appropriate
boundary conditions.
We shall start by investigating the wave functions in the
midi-superspace approximation.
Then the general form of the wave functions and the asymptotic behaviour
of the modes will
suggest the form that the effective interactions take.

\medskip
\noindent{\caps A The Wave Functions}
\smallskip

We shall assume a metric of the form
$$ds^2=\sigma^2\left[(N^2+N_iN^i)d\tau^2+
2N_idx^id\tau+h_{ij}dx^idx^j\right]\mc\eqn$$
where the three-metric $h_{ij}$ is
$$h_{ij}=e^{2\alpha(\tau)}(\Omega_{ij}+\eps_{ij})\ms\eqn$$
Here $\Omega_{ij}$ is the round metric on the unit three-sphere, and we
expand $\eps_{ij}$
in harmonics,
$$\eps_{ij}=\sum_\nlm\bigl[\sqrt6 a_\nlm{1\over3}\QQ\Omega_{ij}+\sqrt6
b_\nlm\Pij+\sqrt2
c_\nlm\Sij+2d_\nlm\Gij\bigr]\mc\eqn$$
with the sum starting at $n=2$, and the notation of [\ref\halh\HALH].
The $n=2$ modes are
rather special: the $n=2$ traceless tensor harmonics all vanish
identically and so the
$n=2$ term in the sum is just the ${\sqrt6\over3} a_2Q^2\Omega_{ij}$
part. (We shall
henceforth drop the labels $\lm$.)
The lapse, shift and scalar field are similarly expanded:
$$\eqalign{N&=N_0\biggl[1+6^{-1/2}\Sn g_n Q^n\biggr]\cr
N_i&=e^\alpha\Sn \bigl[6^{-1/2}k_n\left(P_i\right)^n+\sqrt2
j_n\left(S_i\right)^n\bigr]\cr
\Phi&=\sigma^{-1}\biggl[{1\over\sqrt2 \pi}\phi(t)+\Sn f_n
Q^n\biggr]\ms}\eqn$$
The action is expanded to all orders in $\alpha,\phi, N_0$ and to second
order in the
perturbation quantities $q_n\in(a_n,b_n,c_n,d_n,f_n)$ and
$r_n\in(g_n,k_n,j_n)$. We refer
the reader to [\ref\halh\HALH] for the details of the action,
constraints and three-sphere
harmonics. Conjugate momenta can be defined in the usual manner, and the
hamiltonian
expressed as a function of co-ordinates and momenta. (We denote the
euclidean momenta
conjugate to $q$ by $\pi_q$.) The hamiltonian takes the form
$$H=N_0\biggl[H_{|0}+\Sn H^n_{|2}+\Sn g_n H^n_{|1}\biggr]+\Sn\left(k_n
{}^SH^n_{\_1}+j_n{}^VH^n_{\_1}\right)\mc\eqn$$
with the subscripts denoting the order of each part of the hamiltonian
in the perturbation
quantities, and whether each part arises from varying the lapse or
shift.

We look for wave functions which can be expressed in the form
$$\Psi=\Psi_0(\alpha,\phi)\prod_n \psi_n(\alpha,\phi,q_n)\mc\eqn$$
where $\Psi_0(\alpha,\phi)$ is one of the euclidean WKB solutions of the
mini-superspace
background Wheeler-deWitt equation discussed in Sec.~2 (for instance one
of the
$\psi_k(e^\alpha,\phi)$ defined in \plwaveprod). The full Wheeler-deWitt
equation
$(H_{|0}+\Sn H^n_{|2})\Psi=0$ reduces to a set of euclidean Schrodinger
equations for
$\psi_n$, along the euclidean trajectories defined by the background
wave function. The
partial wave functions $\psi_n$ must also satisfy the linearized
hamiltonian and momentum constraints. The euclidean
trajectories which are defined by the background wave functions
$\psi_k$ are the
solutions to the background classical equations of motion \solution,
with $Q=ik$. We shall consider this choice of background wave function
from now on.
The vertex operator corresponding to this choice
will contain a factor of $e^{ik\phi}$,
from $\psi_k$, and a factor from each of the $\psi_n$,
which will depend on which solution to the euclidean Schrodinger
equation is chosen for
each $n$ mode. This framework, with a WKB background plus quantized
perturbations
satisfying a Schrodinger-type equation, is the `$M$-expansion', which is
discussed further
in [\ref\ll\LL]. We shall later discuss qualitatively the type of wave
functions one might
expect from the modes with $n>2$, but first we consider the $n=2$ modes.
(We drop the mode
label $2$ on the perturbation quantities for ease of notation.)

The partial wave function $\psi_2$ depends just on $\alpha,k,a,f$.
The $n=2$ linear
constraint equations to leading order in the $M$-expansion are
$$\eqalign{{}^S\hat
H_{\_1}\psi_2&={1\over3}e^{-3\alpha}\left(-\hat\pi_a+
a\pi_\alpha+3f\pi_\phi\right)\psi_2=0\cr
\hat H_{|1}\psi_2&=\half e^{-3\alpha}
\left(a(\pi^2_\alpha+3\pi_\phi^2 -3e^{4\alpha}) - 2(\pi_\phi\hat\pi_f-
\pi_\alpha\hat\pi_a)
\right)\psi_2=0\ms}\eqnl\lce$$
The carets on $\pi_a,\pi_f$ denote the operator form of the euclidean
momenta conjugate to
$a,f$ (represented as $-\partial/\partial a,-\partial/\partial f$),
while the quantities
$\pi_\alpha$ and $\pi_\phi$ are the euclidean momenta of $\alpha,\phi$
in the background solution. These are given by
$\pi_\alpha=\sqrt{e^{4\alpha}-k^2}$ and $\pi_\phi=ik$.
We can use the first of equations \lce\ to substitute for
$\hat\pi_a\psi_2$ into the
second equation, which we can then solve on the surface $a=0$. This can
then be used as an
initial condition for the first equation, to obtain
$\psi_2(\alpha,k,a,f)$ at non-zero
values of $a$. We see that $\psi_2$ has the form of a Gaussian at $a=0$,
$$\psi_2(\alpha,k,0,f)=c_2(\alpha,k)
e^{-{3\over2}\pi_\alpha(\alpha,k)f^2}\mc\eqn$$
although it is not clear what $\psi_2$ represents since both $a$ and $f$
are gauge
quantities. The main point is that the linearized constraint equations
have picked out one
solution to the euclidean Schrodinger equation for these modes, which
appears to be in a kind of `ground state'.

For $n>2$ the situation is completely different from the above. In this
case there are
three linearized constraint equations for five perturbation quantities,
so there are two
physical degrees of freedom remaining; one scalar (say $s_n$,
originating from
$a_n,b_n,f_n$) and one tensor ($d_n$).
We can perform a reduction to the physical modes before quantization
(see [\ref\wada\WADA,\ref\shw\SHW]).
Then each of the $n>2$ partial wave functions
$\psi_n$ depends on $\alpha,k,s_n,d_n$ with the $s_n$ and $d_n$ parts
being separable.
Each part obeys a euclidean Schrodinger equation of the form
$$N_0\,{}^{ph}\hat H_{|2}^n\psi_n=-
{\partial\over\partial\tau}\psi_n\mc\eqn$$
where the operator ${}^{ph}\hat H_{|2}^n$ is a diagonalizable
homogeneous quadratic in
$q_n$ and $\hat\pi_{q_n}$, and $q_n\in(s_n,d_n)$. The problem thus
reduces to the euclidean Schrodinger equation for a harmonic
oscillator with time-dependent frequency. (For the tensor
modes $d_n$ in pure gravity this equation reduces to the simple harmonic
oscillator, as shown in [\ref\hfdl\HFDL].) The method of Salusti and
Zirilli [\ref\sz\SZ]
(also discussed in [\ref\berg\BERG]) can be used to find a complete set
of orthonormal
solutions. We describe how this can be done, and the inner product
defined in the euclidean sector, in the Appendix.
The states can be obtained from
a Gaussian ground state by the application of suitable `raising
operators', in a similar
manner to the simple harmonic oscillator.  The action of $m$
raising operators
on the ground state yields a state which is interpreted as describing a
closed universe
containing $m$ particles in that mode. It takes the form
$$\psi_{nm}(\tau,q_n)=\gamma_{nm}(\tau)H_m(\beta_n(\tau)q_n) e^{-
\alpha_n(\tau)q_n^2}\mc\eqnl\nmwave$$ where $H_m$ is an $m$th order
Hermite polynomial.
The solutions to the euclidean Schrodinger equation along the
classical trajectories of fixed $k$ can then be `lifted' to full
solutions of the
Wheeler-deWitt equation, using the method described in $[\ref\ll\LL]$.

\medskip
\noindent{\caps B The Classical Solutions and the Vertex Operators}
\smallskip

There are two ingredients in the path integral formula for the matrix
element \matrix. One
is the wave function, which describes the quantum state of the wormhole.
We have described
above the general form that such wave functions take. The other is the
path integral over
asymptotically flat four-geometries with an inner boundary on which the
three-geometry and
matter field are specified. For the semi-classical evaluation of this
path integral we
need the classical euclidean solutions for the perturbation modes. We
shall now discuss these.

The classical euclidean solutions including the $n=2$ modes can be
obtained from the
background solution \solution\ by slicing the background in a different
way. The
background solution can be expressed in the form
($\infty>r>r_0>\sqrt{|Q|/2}$)
$$\eqalign{ds^2&=\sigma^2\Omega(r^2)(dr^2+r^2d\Omega_3^2)\cr
\phi&=\phi(r^2)\mc}\eqn$$
where $$\eqalign{\Omega(X)&=1-{1\over4}Q^2X^{-2}\cr
\phi(X)&=\phi_-+\half\ln\left[{X+\half
Q\over X-\half Q}\right]\ms}\eqn$$
We now change co-ordinates from $r,\chi$ to $r',\chi'$, with
$r\cos\chi=r'\cos\chi'-\eps$
and $r\sin\chi=r'\sin\chi'$ (so that $r^2=r'^2-2\eps r' \cos\chi'
+\eps^2$), for a
constant parameter $\eps$. This change is an isometry of the flat metric
$dr^2+r^2d\Omega_3^2$. However $\Omega$ and $\phi$ develop
non-spherically symmetric parts
in the new slicing of ${\bf R}\times S^3$ in three-spheres of constant
$r'$. (We also
change the boundary to one at constant $r'$, instead of at constant
$r$.) We see that
$$\eqalign{\Omega(r^2)&=\Omega(r'^2)-\eps Q^2 r'^{-5} \cos\chi' +
O(\eps^2)\cr
\phi(r^2)&=\phi(r'^2)+\eps Qr'^{-3}\Omega^{-1}(r'^2)\cos\chi'+
O(\eps^2)\ms}\eqnl\falloff$$
as $\eps\tendsto0$. We thereby obtain an exact solution to the
linearized equations for
the $n=2$ modes, since $\sqrt{2/\pi}\cos\chi$ is an $n=2$ scalar
harmonic. We shall now
remove the primes on $r',\chi'$. (Notice that in this form the solution
is obtained in the
particular gauge $k=0,\ g=a,\ N_0=e^\alpha r^{-1}$.)

The $n=2$ scalar field perturbation falls
off like $r^{-3}\cos\chi$ in the asymptotic region ($r\tendsto\infty$),
while the
background part of the scalar field falls off like $r^{-2}$. It is
essentially the fact that $\phi$ falls off with proper distance
like the two-point function in flat space which
led to the vertex operator being a function of $\phi$
(no derivatives of $\phi$) for the spherically
symmetric model of Sec.~2.
We see from \falloff\ that the $n=2$ mode falls off like a single
derivative of the
two-point function $c^\mu\partial_\mu r^{-2}=-2r^{-3}\cos\chi$, where
$c^\mu$ is a
constant vector in flat space. Thus the vertex operator coming
from these modes
would be expected to be a function of $\partial_\mu\phi$.  By
considering the asymptotic
form of the linearized equations for the general $n$ mode, we see that
the scalar part
falls off like $r^{-n-1}Q_n$, for a scalar harmonic $Q_n$. Thus the
contribution from the
$n$th mode will produce a vertex operator which is made up of $n-1$
derivatives of $\phi$.
There is an integration over the $O(4)$ rotations at the end of the
calculation (a zero
mode integration, similar to the integral over $x_0$ in Sec.~2). This
means that the total vertex operator will be Lorentz covariant.
In general, a wave function in the scalar sector of the form \nmwave\
would be expected to
lead to a vertex operator $(V_n)^m$ which is the $m$th power of a basic
operator $V_n$.
The operator $V_n$ will be made up of $n-1$ derivatives of $\phi$.
However, we have found for the $n=2$ mode, that the wave function
satisfying the linearized hamiltonian and momentum constraints is forced
to have Gaussian
form. For such wave functions the vertex operator will be restricted to
have $m=0$. Thus
we have appeared to rule out the operators $(\partial\phi)^m$, for
$m>0$, but not
corresponding operators involving higher derivatives of $\phi$. This
appears to contradict
work in [\ref\hfdthesis\HFDTHESIS], which considers the conformally
coupled massless
scalar field $n=2$ mode. In [\ref\hfdthesis\HFDTHESIS] it appears that
the vertex operator
$(\partial\phi)^2$ is present, corresponding to the lowest allowed
excited solutions to
the Wheeler-deWitt equation. We expect however, that the conformally and
minimally coupled
massless scalar fields should have the same number of degrees of freedom
for each level
$n$, and that our result on the absence of first derivative terms in the
vertex operator
would apply also in the conformally coupled case.

Our argument fixing the form of the $n=2$ wave function to be in its
ground state relied
on the fact the $n=2$ modes are pure gauge. It does not apply to the
modes $n>2$. For
the $n>2$ modes excited states will exist which satisfy the
Wheeler-deWitt equation and
linearized constraints. It is expected that the part of the vertex
operator (in the scalar
part of the theory) coming from the $m$th excited state will consist of
$m$ powers of
operators involving $n-1$ derivatives of $\phi$. This will yield
corrections to the vertex
operator $e^{ik\phi}$ in the massless scalar field model which cannot be
obtained using
the Green functions of fields in the spherically symmetric wormhole
background
[\ref\gms\GMS]. Thus the issue of whether the semi-classical vertex
operator has really
been found to be $e^{ik\phi}$ to all orders in the wormhole scale, is
still a matter of debate.

\newsection{4 Conclusions}

We have discussed the form of the vertex operator for wormholes in the
massless minimally
coupled scalar field model. We used the wave function approach of
Hawking
[\ref\swhpr\SWHPR], and described the evaluation of vertex operators
corresponding to a
basis of solutions to the hamiltonian and momentum constraints. The
$O(4)$-symmetric
mini-superspace truncation yielded a correspondence between a particular
basis of wave
functions which separated in the $(a,\phi)$ mini-superspace variables,
and effective
semi-classical vertex operators of the form $\int d^4x_0 \alpha(k)\colon
e^{ik\phi(x_0)}\colon$. These vertex operators had previously been
obtained by considering
the field theory in the background of an imaginary scalar field
euclidean wormhole
solution [\ref\gm\GM]. Using that method it appears that there are no
further corrections
to the vertex operator of higher order in $l_{pl}$ (or the wormhole
characteristic scale).
However, using instead the wave function approach it appears that such
corrections exist,
and would come from non-spherically symmetric perturbation modes on the
spherically
symmetric background. The $n=2$ modes, which would yield a factor in the
vertex operator
like $(\partial\phi)^m$, appear to be pure gauge, and so the linearized
hamiltonian and
momentum constraints restrict this part of the wave function to be in
its ground state.
Thus these terms appear not to be present. This appears to contradict
earlier work on the
conformally coupled scalar field model [\ref\hfdthesis\HFDTHESIS].
However the $n>2$ modes
can be excited, and would appear to lead to higher derivatives in the
effective
interaction. This contrasts with the results obtained in [\ref\gms\GMS]
using the field
theory in the imaginary scalar field wormhole background, and suggests
that the two
different approaches to wormholes correspond to physically different
situations.

\newsection{Acknowledgements}

I am grateful to D. Page, R. Laflamme, A. Barvinski, J. Louko
and J. Twamley  for interesting and
informative discussions relating to this work. This research has been
supported by NSERC (Canada) and the Winspear Foundation.

\appendix{A}
\vskip .5cm \goodbreak \centerline{\bf Appendix}
\nobreak\vskip .5cm\nobreak\eqnnumber=0 \subsectionno=1

We show here how to construct the basis of solutions to the
euclidean Schrodinger equation which are used in Sec. 3. We also
construct an inner product which is particularly natural for wormholes,
and with respect to which the wave functions are orthonormal.

Consider the case where the euclidean WKB background wave functions
are the $\psi_k(e^\alpha,\phi)$ defined in \plwaveprod.
Each wave function $\psi_k$ defines
a euclidean Hamilton-Jacobi function
$$I_k(\alpha,\phi)=\half\left(\sqrt{e^{4\alpha}-k^2}+k\arcsin
(ke^{-2\alpha})\right)+ik\phi\ms\eqn$$
We shall denote the partial derivatives by
$I_\alpha=\sqrt{e^{4\alpha}-k^2}$
and $I_\phi=ik$.
There are associated euclidean trajectories, which are the solutions
to the equation of motion \solution, with $Q=ik$. The classical
euclidean momenta are given by $\pi_\alpha=I_\alpha,\ \pi_\phi=
I_\phi$. Since the background wave functions have
WKB form, in the region $e^{4\alpha}\gg k^2\gg 1$, we can reduce
the quantization of the $n>2$ perturbations to the study of the
Schrodinger equation for one tensor and one scalar mode, in the
background solution [\ref\shw\SHW]. These are the physical degrees
of freedom, once all linearized constraints have been solved.
In our case, the background classical geometries are euclidean,
and one obtains a {\it euclidean} Schrodinger, or heat, equation.
The equation is generally of the form
$$H_n\psi_k(\tau,q_n)=-{1\over N_0}{\partial\over\partial\tau}
\psi_k(\tau,q_n)\mc\eqnl\esch$$
where the `time parameters' $\tau$ along the trajectories have been
introduced via $\dot\phi e^{3\alpha} = N_0 I_\phi$ and
$-\dot\alpha e^{3\alpha} = N_0 I_\alpha$.
This can be conveniently solved in the gauge $N_0=e^{-\alpha}$ by
$\tau=-\half I_\alpha$.

Leaving out the suffices $n$, and choosing a
convenient operator ordering, the `hamiltonian' $H$ is
$$H=\half e^{-3\alpha}\left( A\partial^2_q +\half B
(q\partial_q+\partial_q q)
+Cq^2\right)\mc\eqn$$
where, for the tensor modes ($q=d$), $A=-1,\ B=0,\
C=e^{4\alpha} (n^2-1)$, and for the scalar modes ($q=s=
f+{I_\phi\over I_\alpha}(a+b)$),
$$\eqalign{A&=-\left(1+{3\over n^2-4}
\left(I_\phi\over I_\alpha\right)^2\right)\cr
B&=6\left(n^2-1\over n^2-4\right){I_\phi^2\over I_\alpha}\cr
C&=(n^2-1)\left(I_\alpha^2-{n^2+5\over n^2-4}I_\phi^2\right)
\ms}\eqn$$
The same procedure as used in [\ref\shw\SHW] has been used to
obtain these equations, except for the difference that this
analysis is in euclidean time.

By writing
$$\psi_k(\tau,q) = e^{-{B\over4A}q^2}\,
\chi_k(\tau,q)\mc\eqnl\chitopsi$$
the equation for $\chi_k$ takes the form of \esch\ with
$A\tendsto A,\ B\tendsto 0$ and
$$C\tendsto C'=C-{B^2\over4A}-{e^{-3\alpha}\over2N_0}{d\over d\tau}
\left(B\over A\right)\ms\eqn$$
Solutions exist for $\chi_k$
which are of Gaussian form
$$\chi_k(\tau,q)=u^{-1/2}\exp\left({\dot u e^{3\alpha}
\over 2uN_0A}q^2\right)\mc\eqn$$
for any $u$ satisfying the linear equation
$${1\over N_0}{d\over d\tau}\left(\dot ue^{3\alpha}\over N_0 A\right)
+ C'e^{-3\alpha} u = 0 \ms\eqnl\ueqn$$
The equation
\ueqn\ has a number of important properties. Firstly, it is real
and so it possesses two linearly independent
real solutions. This is not the case for its lorentzian counterpart.
Secondly, in the asymptotic region, $\tau\tendsto -\infty$,
$A\tendsto -1$ and $C'\sim e^{4\alpha}(n^2-1)$, so
\ueqn\ reduces to the equation for the tensor modes.
One of the solutions to \ueqn\ tends to zero
while the other tends to infinity, in this region.
We can use this to define a preferred vacuum wave
function, $\chi_{k0}(\tau,q)$ which is
obtained by choosing the solution to \ueqn,
$u(\tau)=u_0(\tau)$, which tends to zero in the asymptotic region.
(In the gauge $N_0=1$ we have
$u_0(\tau)\sim\tau^{-n-1}$.) Thirdly, we note that \ueqn\ is
unchanged with respect to $\tau\tendsto-\tau$.
(If we choose, for instance, the gauge $N_0=e^{-\alpha}$,
with the choice
of time parameter $\tau=-\half I_\alpha$, this is made somewhat
more explicit.)
The origin of this symmetry is the discrete isometry of the
wormhole geometry, which interchanges the two asymptotic regions.
This symmetry means that if $u(\tau)$ is
a solution to \ueqn, then so is $\tilde u(\tau)=u(-\tau)$.

The `vacuum' wave function, $\chi_{k0}$, is annihilated
by the operator
$$a(u) = u\partial_q - {\dot u e^{3\alpha}\over N_0 A} q\mc\eqn$$
with $u=u_0$.
The operator $a(\tilde u_0)$, when applied to $\chi_{k0}$, yields
another solution to the euclidean Schrodinger equation,
(non-vanishing, because $u_0(\tau)$ and $\tilde u_0(\tau)$
are linearly independent).
We call this solution $\chi_{k1}$.
This process can be continued indefinitely to
produce $\chi_{km}$.
The operator $a(\tilde u_0)$ is the `generalized raising operator'.
We can then construct $\psi_{km}$ from $\chi_{km}$ using \chitopsi.

To construct the appropriate inner product, in
which these wave functions are all orthogonal, we first use
$\tilde u_0(\tau)$ to define a `dual vacuum'
$$\eqalign{\tilde \chi_{k0}(\tau,q)&=
\tilde u_0^{-1/2}\exp\left(-{\dot{\tilde u}_0 e^{3\alpha}
\over 2\tilde u_0N_0A}q^2\right)\cr
\tilde\psi_{k0}(\tau,q)&=
e^{+{B\over4A}q^2}\tilde\chi_{k0}(\tau,q)\mc}\eqn$$
which satisfies the `time reversed' euclidean Schrodinger equation
$$
\half e^{-3\alpha}\left(A \partial_q^2
-\half B (q\partial_q+\partial_q q)+ Cq^2\right)
\tilde\psi_k(\tau,q) = +{1\over N_0}{\partial\over\partial\tau}
\tilde\psi_k(\tau,q)\ms\eqn
$$
(Under time-reversal we note that $B\tendsto -B$.)
The operator
$$\tilde a(u) = -u\partial_q - {\dot u e^{3\alpha}\over N_0 A} q
\mc\eqn$$  annihilates $\tilde\chi_{k0}$, if $u=\tilde u_0$, and
if $u=u_0$ it can be applied $m$ times to $\tilde\chi_{k0}$, to
produce $\tilde\chi_{km}$ and hence the `dual $m$th excited state',
$\tilde\psi_{km}$.

With these definitions, the inner product
$$(\phi,\psi) = \int_{-\infty}^{+\infty} dq \tilde\phi(\tau,q)
\psi(\tau,q)\eqn$$
is independent of $\tau$, and with suitable normalization constants
the wave functions $\psi_{km}$ are orthonormal with respect to this
inner product.

The wave functions $\psi_{km}(\tau,q_n)$ have the form quoted in the
text,
$$\psi_{km}(\tau,q_n)=\gamma_{knm}(\tau)H_m(\beta_{kn}(\tau)q_n)
e^{-\alpha_{kn}(\tau)q_n^2}\mc\eqnl\nmwave$$
where $H_m$ is an $m$th order Hermite polynomial.
They are interpreted in terms of quantum states corresponding to a
closed universe containing $m$ particles in the $n$th harmonic mode.

\newsection{References}

{\obeylines
\item {[1]}\gpope
\item {[2]}\colcc
\item {[3]}\swhpr
\item {[4]}\hp
\item {[5]}\gms
\item {[6]}\arlo
\item {[7]}\hfd
\item {[8]}\hfdl
\item {[9]}\arlnew
\item {[10]}\halh
\item {[11]}\hfdthesis
\item {[12]}\coll
\item {[13]}\gsax
\item {[14]}\jjhjl
\item {[15]}\lee
\item {[16]}\gm
\item {[17]}\bateman
\item {[18]}\hh
\item {[19]}\arlwip
\item {[20]}\ll
\item {[21]}\wada
\item {[22]}\shw
\item {[23]}\sz
\item {[24]}\berg}
\bye